\documentstyle[aps,preprint,prl,epsfig]{revtex}

\begin{document}


\title
{
Large-scale simulation of adhesion dynamics for end-graphed polymers
}

\date{\today}

\author{
Scott W. Sides,
Gary S. Grest,
and Mark J. Stevens
}

\address
{
Sandia National Laboratories,
Albuquerque, New Mexico 87185-1411
}

\maketitle

\begin{abstract}

The adhesion between a polymer melt and substrate
is studied in the presence of chemically attached chains on the substrate surface.
Extensive molecular dynamics simulations have been carried out to
study the effect of temperature,
tethered chain areal density ($\Sigma$), tethered chain length ($N_{\rm t}$),
chain bending energy ($k_{\theta}$)
and tensile pull velocity ($v$) on the adhesive failure mechanisms
of pullout and/or scission of the tethered chains.
We observe a crossover from pure chain pullout to chain scission
as $N_{\rm t}$ is increased.
Below the glass transition, the value of $N_{\rm t}$ for which this crossover begins
approaches the bulk entanglement length $N_{\rm e}$.
For the values of $N_{\rm t}$ and $\Sigma$ used here, no crossover to crazing
is observed.

\end{abstract}



\section{Introduction}

Adhesion at polymer interfaces is important in many diverse applications
such as, colloidal stabilization, \cite{napper83_book,russell89_book}
filler modification of polymeric materials and lubrication \cite{jklein96}
and enhancing mechanical properties of polymer blends. \cite{brown93_2,riess_review75}
The interface of interest can be either:
(i)  between two homopolymer melts (A+B) or
(ii) between a homopolymer melt and a hard surface (A+substrate).
Most polymer blends do not mix, because even a weak repulsive interaction causes
the system to phase separate as the chain length increases.
This phase separation reduces entanglements at the interface and
the adhesive strength is then due to relatively weak van der Waals forces alone.
For case (ii), the interfacial strength depends on the number of
chemical bonds between monomers of the polymer and sites on the substrate.
In the absence of strong chemical bonding, the adhesive strength is again
dominated by weak van der Waals attractions.
For both cases certain additives or compatibilizers can increase
the adhesive strength of the interfaces.
For case (i), the additive can be an AB diblock copolymer.
The A(B) block of the copolymer can form entanglements with the A(B) melt side
respectively.
In case (ii), the additive is a chain of the A species with a functionalized
end-group able to react with the substrate, thereby forming a strong chemical bond.
The tethered chain of the A species can then become entangled with the melt.
In both cases, larger numbers of polymer entanglements result in
increased interfacial strength.
The systems simulated in this paper consist of substrate-tethered chains in contact
with a polymer melt (ie case (ii)), but the results can easily be generalized to
polymer/polymer interfaces.

Two key parameters govern the amount of interfacial entanglement,
the length ($N_{\rm t}$) and the areal density ($\Sigma$) of the tethered chains.
Adhesion enhancement due to a tethered polymer layer shows a surprising
non-monotonic behavior as a function of $N_{\rm t}$ and $\Sigma$ due
to the phase behavior for the tethered chain layer.
\cite{deGennes80_brush,aubouy95,grest96_brushes,leger_review99}
The increased work of adhesion $W^*$ for an elastomeric system
versus areal density of
graphed chains $\Sigma$ is shown schematically in Fig.~\ref{fig_sch_WvsA}.
$W^*$ is defined as the extra work needed to separate two surfaces in
addition to the work exerted against dispersion forces alone.
In Region I where each tethered chain is isolated from its neighbors,
the work of adhesion increases as $\Sigma$ increases.
However as $\Sigma$ continues to increase, the tethered chains begin
to overlap and phase separate from the melt (region II).
In this regime, the interpenetration of the tethered chains into the polymer
melt decreases with increasing $\Sigma$, and as a result $W^*$ decreases.
At sufficiently high areal densities, the polymer melt is completely expelled
from the tethered chains thereby causing the work of
adhesion to fall off to the bare value due to dispersion forces only (region III).
$W^*$ has a similar dependence on $N_{\rm t}$.
This non-monotonic behavior has been observed by
L{\' e}ger and co-workers \cite{leger_review99,leger_tirrell95} in elastomeric materials.
Kramer and co-workers
\cite{kramer91,wash_kramer94,kramer94_adhesion_review,kramer95_tether,kramer96_copoly}
have studied the effects of tethered chains in glassy polymer systems
where the effects of crazing could complicate the dependence of $W^*$ on $\Sigma$.
While Fig.~\ref{fig_sch_WvsA} relates the macroscopic work
of adhesion to tethered-chain parameters, it does
not address the {\it microscopic} processes involved in adhesion failure at the
melt/tethered-chain interface.
This paper investigates the effects of $N_{\rm t}$, $\Sigma$, $T$, $k_{\theta}$ and $v$
on these failure mechanisms.

The interplay between the microscopic failure mechanisms of
tethered chain pullout and scission (including their possible relationship to crazing)
are not fully understood, partly due to the difficulty
of direct experimental observation of these phenomena.
Molecular dynamics (MD) simulations of fracture in
highly crosslinked systems \cite{stevens_fracture1_00,stevens_fracture2_00}
and crazing \cite{robbins01,robbins99}
have helped to elucidate the crossover
from adhesive to cohesive failure of polymer adhesives
near walls {\it without} end-tethered chains.
Simulations of tethered chains on small, highly simplified
models in 2D have been performed, \cite{balazs96,grant99}
but were unable to study the effects of chain scission in particular.
This paper extends the results on our earlier work \cite{sides01_ad1}
of large-scale simulations to study
the adhesive failure mechanisms of end-tethered chains in contact
with an entangled polymer melt.
Section II contains details related to the MD simulations and the method
used to implement breakable bonds.
In Sec.~III results of the MD simulations are presented and conclusions
are given in Sec.~IV.

\section{MD model/method}

We perform continuous-space, molecular dynamics (MD) simulations
on a coarse-grained model of polymer chains.
The polymers are represented by attaching $N_{\rm t}$ spherical
beads of mass $m$ with breakable springs.
Bead trajectories are obtained by stepwise integration
of Newton's equations of motion (EOM)
using a velocity-Verlet \cite{tild87} algorithm with a time step
$\Delta t$=$0.006 \tau$, where
$\tau$=$\sigma(m / \epsilon)^{1/2}$, 
with $\sigma$ and $\epsilon$
setting the length and energy scales respectively.
The EOM includes terms for
a weak stochastic force and a viscous drag force with a
coefficient on the viscous force term of $0.5 \tau^{-1}$.
\cite{kremer_grest90}
The addition of these two forces to the EOM effectively
couples the system to a heat bath.
Each pair of monomers separated by a distance $r$ interacts through
a standard (12-6) Lennard-Jones potential $U_{\rm LJ}(r)$,
 \begin{eqnarray}
 \label{eq_lj_potential}
 \nonumber
  U_{\rm LJ}(r)
   =
   \left \{ \begin{array}{ll}
    4 \epsilon \left [ \left (\frac{\sigma}{r} \right )^{12} -
                       \left (\frac{\sigma}{r} \right )^{6}
               \right ]  & r < r_{c} \\
    0                    & r > r_{c},
   \end{array} \right.
 \end{eqnarray}
with $r_c$=$2.2 \sigma$.
For simplicity, the substrate interaction is modeled as a
flat wall by an integrated LJ potential
 \begin{eqnarray}
 \label{eq_ljwall_potential}
 \nonumber
  U_{\rm LJ}^{\rm wall}(z)
   =
   \left \{ \begin{array}{ll}
    \frac{2 \pi \epsilon_{\rm wall} }{3}
    \left [ \frac{2}{15} \left (\frac{\sigma}{z} \right )^{9} -
                         \left (\frac{\sigma}{z} \right )^{3}
               \right ]  & z < z_{c}^{\rm wall} \\
    0                    & z > z_{c}^{\rm wall},
   \end{array} \right.
 \end{eqnarray}
with $z_c^{\rm wall}$=$2.2 \sigma$,
$\epsilon_{\rm wall}$=$2.0 \epsilon$ (strongly attractive wall, no tethered chains) and
$\epsilon_{\rm wall}$=$0.1 \epsilon$ (weakly   attractive wall, tethered chains).

To study the effect of chain scission on adhesion, the
standard finite extensible non-linear elastic
(FENE) potential \cite{kremer_grest90}
is altered to allow for broken bonds along a polymer chain.
The total non-breakable potential $U_{\rm nb}(r)$
(including the FENE potential),
between two adjacent beads on the same chain separated by
a distance $r$ takes the form
 \begin{eqnarray}
 \label{eq_fene_potential}
 \nonumber
  U_{\rm nb}(r)
   =
   \left \{ \begin{array}{ll}
     U_{\rm LJ}(r) +
     -0.5 K R_{o}^2 \ln  \left [ 1 - \left (\frac{r}{R_o} \right)^2 \right ]
     & r < R_{o} \\
     \infty
     & r > R_{o}
   \end{array} \right .
 \end{eqnarray}
while the breakable potential $U_{\rm b}(r)$, takes the form
 \begin{eqnarray}
 \label{eq_break_potential}
 \nonumber
  U_{\rm b}(r)
   =
   \left \{ \begin{array}{ll}
     U_{\rm LJ}(r) +
     k r^4 \left [ ( r - r_1 ) ( r - r_2 ) \right ] +
     U_c
     & r < r_{\rm br} \\
     0
     & r > r_{\rm br}
   \end{array} \right .
 \end{eqnarray}
\noindent
with $r_{\rm br}$ the length at which
a chain bond is defined as broken.
MD simulations on a similar coarse-grained model \cite{kremer_grest90}
using the FENE potential provide details of the chain conformations
needed to construct the appropriate starting states.
The parameters in the bond-breaking potential
come from fitting $U_{\rm b}(r)$ to the region of the aforementioned
FENE potential
($K$=$30.0 \epsilon / \sigma^{2}$ and $R_o$=$1.25 \sigma$)
near its mean equilibrium bond length.
This fitting procedure results in 
$r_{\rm br}$=$1.21 \sigma$ and a bond-breaking barrier
$\Delta U_{\rm b} \approx 20 \epsilon$.
The remaining parameters in $U_{\rm b}(r)$ are
$k$=$-409.12 \epsilon / \sigma^{6}$,
$r_1$=$1.2   \sigma$,
$r_2$=$1.219 \sigma$ and
$U_c$=$42.059 \epsilon$.
Fitting $U_{\rm b}(r)$ to a FENE potential this way
allows constructing an initial configuration with chains of
approximately the correct radius of gyration $R_g$ by utilizing
the aforementioned
chain-building algorithm. \cite{kremer_grest90}
Figure \ref{fig_ubreak} compares the standard non-breakable
potential to the breakable potential used in this study.
The form of $U_{\rm b}(r)$ allows for two extrema in the bond potential;
one stable, global minimum near the FENE
potential energy minimum, and one local maximum near $r_{\rm br}$
where the bond force becomes zero.
This allows bonds to be removed safely from the force calculation
without causing large recoil velocities on the resulting chain ends.

The model outlined above results in completely flexible chains.
Experimentally, real polymer chains are not completely flexible but 
have some degree of stiffness.
Chain stiffness is added to the model by adding a bending potential energy
for each triplet of consecutive particles along a chain.
The bending potential takes the form \cite{faller00}
 \begin{eqnarray}
 \label{eq_faller_potential}
 \nonumber
  U_{\theta}(\theta) & = &
   k_{\theta} \left ( 1 + \cos \theta \right )
 \end{eqnarray}
\noindent
with an equilibrium angle
of $\theta_0$=$\pi$ where $\theta$ is defined in Fig.~\ref{fig_sch_bend}.
When a bond between two monomers is broken during a simulation, all of the triplets
associated with that bond are removed from the force calculation.

The initial chain configurations must be constructed appropriately
for the model and potential parameters chosen.
For the systems of very long, entangled chains used in this study,
it is not feasible to equilibrate the chains by brute-force MD simulation.
In general, it is not known how to construct the correct configurations for
end-tethered chains next to a long-chain melt.
To avoid difficulties in initializing systems
for the various wet brush regimes
we restrict the present study to areal densities $\Sigma$ in
the so-called ``mushroom regime'' \cite{deGennes80_brush,aubouy95}
(i.e.~Region I in Fig.~\ref{fig_sch_WvsA}).
In this regime each tethered chain interacts weakly
with other tethered chains and may be constructed as a Gaussian chain.
So, for a given $N_{\rm t}$ and $\Sigma$,
all chains are constructed as random walks with the
correct mean value of $R_g$.
The tethered chains are attached to the substrate wall in random locations.
The correct value of $R_g$ depends on $k_{\theta}$, which
is known from recent bulk simulations. \cite{faller00_thesis}
The system size is adjusted so that the tethered chains
do not bridge the box, thereby interacting with the opposite wall.
A soft potential is used initially to remove overlaps prior to switching
on the full LJ potential between monomers.
The size of the simulation cell is adjusted until the pressure $P \approx 0$
resulting in an overall monomer density of
$\rho \approx 0.85 \sigma^{-3}$
($\rho \approx 0.88 \sigma^{-3}$)
for the
highest
(lowest) temperatures used.
For all simulations, the number of beads in each of
the melt chains is $N_m$=$2500$
and the number of tethered chains is $n_{\rm t}$=$30$.
The chain stiffness is taken to be either $k_{\theta}$=$0$ (fully flexible),
or $k_{\theta}$=$1.5 \epsilon$
for which the entanglement length as calculated from the plateau modulus
is $N_{\rm e}$=$72$ and $27$ monomer units respectively. \cite{putz00,faller01}
The temperatures used range from
$T$=$1.0 \epsilon/k_{\rm B}$,
which is well above the glass transition temperature
for fully flexible chains
$T_{\rm g}$=$0.5-0.6 \epsilon/k_{\rm B}$,
\cite{robbins01} down to $T$=$0.3 \epsilon/k_{\rm B}$.
Areal densities as low as $\Sigma$=$0.002 \sigma^{-2}$ with $T$=$1.0 \epsilon/k_{\rm B}$ and
$v$=$0.0167 \sigma \tau^{-1}$ were used without any significant differences with
runs using the same parameters for $\Sigma$=$0.008 \sigma^{-2}$.
These results confirm the estimates of $\Sigma$ which place a system in the mushroom regime.
Therefore, to increase computational speed the majority of the simulations
(all the data presented in this paper) use $\Sigma$=$0.008 \sigma^{-2}$.
For systems with crosslinking,
the potential for the crosslinks is the same as the interactions between
monomers bonded along a chain.
The crosslinks are randomly distributed in the system at an average distance along
each chain of approximately $70$ monomers, which is comparable to $N_{\rm e}$.
The crosslinks are added only to the melt chains after the chain configurations
have been constructed and equilibrated.
Simulation are performed using the massively parallel MD code LAMMPS \cite{plimpton95}
(suitably adapted to include chain scission) developed at Sandia and run on the
ASCI Red Teraflop machine and Computational Plant (Cplant) clusters.

Figure \ref{fig_pic_n100} shows chain configurations at different
times from a tensile pull simulation consisting of
approximately $7 \times 10^4$ particles with $N_{\rm t}$=$100$.
Figure \ref{fig_pic_n250} shows chain configurations at different
times from a tensile pull simulation consisting of
approximately $2 \times 10^5$ particles with $N_{\rm t}$=$250$.
Our largest systems contain close to $10^6$ particles.
The tensile pull is achieved by moving only the bottom wall
at constant velocity.
The tethered chains are bonded to the bottom wall of the
simulation cell and the top wall has no tethered chains.
The interaction strength of the top wall is set to be
sufficiently strong so that no adhesive failure occurs on
the top wall during pulling.
The interaction of the bottom wall with the melt has a
very weak attractive component, so that the
adhesion enhancement due to the tethered chains may
be studied independently of the adhesion to the bare wall.
The $z$ axis is normal to both walls
and periodic boundary conditions are used in the
$x$ and $y$ directions.
The size of the simulation cell in the $z$ direction
is set to be approximately three times the radius of
gyration of the tethered chains prior to pulling.
The red chains are tethered, blue chains represent the melt
and green chains were initially tethered and have broken
sometime during the tensile pull.
Figures \ref{fig_pic_n100} and \ref{fig_pic_n250} show the qualitative
dependence of the tethered chain dynamics on $N_{\rm t}$.
For small $N_{\rm t}$, almost all of the tethered chains are completely
pulled out of the melt.
For relatively large $N_{\rm t}$, all of the tethered chains break somewhere
along their length before being totally extracted from the melt.

\section{Results}

During the simulation we measure the total work exerted
and the number of bonds broken by pulling the bottom wall.
If $w(t)$ is the instantaneous work at time $t$
then the total integrated work at $t$ is
 \begin{eqnarray}
  \label{eqn_total_work}
  \nonumber
  {\cal W}(t) & = & \int_0^t w(t') dt'
 \end{eqnarray}
To quantify the degree of scission
the remaining length of the tethered chains is averaged
at the end of a pull simulation.
The average fractional length of the remaining chains is
 \begin{eqnarray}
  \label{eqn_F}
  \nonumber
  \langle F \rangle = \frac{1}{n_{\rm t}}
                      \sum^{n_{\rm t}}_i \frac{N_{{\rm t},i}^f}{N_{\rm t}} \
 \end{eqnarray}
where $N_{\rm t,i}^{f}$ is defined to be length of the $i$th
tethered chain at the end of a pull simulation and
$n_{\rm t}$ is the number of chains tethered to the pulling wall.
The value of $\langle F \rangle$ quantifies the amount of
scission in the tethered chains at the end of a pull.
$\langle F \rangle$=$1$ corresponds to pure chain pullout and
$\langle F \rangle$=$0$ to chain scission at the wall for each chain.

\subsection{Chain length dependence}

Figure \ref{fig_Wvst_T03_Ns}
shows the integrated work over time as a function of time ${\cal W}(t)$,
at different values of $N_{\rm t}$ for
$T$=$0.3 \epsilon / k_{\rm B}$, which is well below $T_{\rm g}$.
For each tethered chain length, ${\cal W}(t)$ is measured for a
sufficiently long time
such that each of the tethered chains have either completely
pulled out or broken.
The plateau in ${\cal W}(t)$ at large $t$ signifies the complete
debonding of the pulling surface from the entangled melt.
Surfaces with longer tethered chains require more work
to completely pull away from the polymer melt than do short chains.
However, the maximum work required for surface debonding saturates
for large $N_{\rm t}$.
This effect can be explained by examining the chain length dependence on
tethered chain scission.

Figure \ref{fig_FvsN_Vs}a shows~$\langle F \rangle$ versus
$N_{\rm t}$ for two values of $v$ at a temperature well above
the glass transition.
For $N_{\rm t} \lesssim 50$ the chains completely pull out of the melt
with no breaking.
For $N_{\rm t} \gtrsim 50$ only a fraction of each tethered chain pulls out
of the melt before breaking.
For the longer tethered chain lengths the value of~$\langle F \rangle$
first decreases and then saturates, consistent with the data for ${\cal W}(t)$ shown
in Fig.~\ref{fig_Wvst_T03_Ns}.
This data shows the entanglement length of the tethered chains
to be an important length scale.
For the fully flexible model ($k_{\theta}$=$0.0$),
$N_{\rm e} \cong 72$ which is consistent with a value between $50$ and $100$
for the location of the crossover to chain scission as
seen in Fig.~\ref{fig_FvsN_Vs}a.
Figure \ref{fig_FvsN_Vs}b shows~$\langle F \rangle$ versus
$N_{\rm t}$ for two values of $v$ for $T$ well below $T_{\rm g}$.
This data has the same qualitative behavior on $N_{\rm t}$ as the data
in Fig.~\ref{fig_FvsN_Vs}a, however it is less dependent on the pulling
velocity $v$ and the crossover to scission occurs at a tethered
chain length close to $N_{\rm e}$ for both values of $v$.
The normal mode relaxation time of a chain in an entangled polymer
melt increases as the wavelength of the mode increases. \cite{kremer_grest_review}
Therefore as the pulling speed $v$ becomes slower, a larger average fraction
of each tethered chain is able to pull out of the melt before breaking.
As $T$ is lowered below the glass transition however,
the dynamics are dominated by the increased monomeric friction and the velocity
dependence on chain scission is considerably weaker.

This is more clearly illustrated in Fig.~\ref{fig_FvsV_Ts} which shows
$\langle F \rangle$ dependence over two decades in velocity $T$
above and below $T_{\rm g}$.
For $T < T_{\rm g}$, the data is nearly independent of the pulling speed.
For $T > T_{\rm g}$, Fig.~\ref{fig_FvsV_Ts} includes data for an uncrosslinked
melt as well as a crosslinked network.
In most applications, polymer adhesives are crosslinked when $T$ is
above $T_{\rm g}$ in order to improve mechanical properties.
For large $v$, the data show that the amount of chain scission depends
little on whether or not the polymer melt is crosslinked because the melt
chains cannot relax over short times.
This weak dependence of $\langle F \rangle$ on crosslinking at large $v$
can also be seen in Fig.~\ref{fig_FvsN_Vs}a (open squares).
Presumably, if $v$ is decreased further in the uncrosslinked system for $T > T_{\rm g}$
then the chains would pullout completely of the polymer liquid.
However for the crosslinked network at $T > T_{\rm g}$, the value of $\langle F \rangle$
should begin to saturate for the lowest pulling velocity, albeit with less chain scission
than for all the systems below $T_{\rm g}$.
Extensive simulations on entangled melts \cite{putz00} have determined that the
entanglement time $\tau_{\rm e} \approx 1400 \tau$, which is the average time needed for
a chain to crossover from Rouse-like to reptation-like dynamics.
The smallest values of $v$ in Fig.~\ref{fig_FvsV_Ts} correspond to pullout
times (i.e.~the time required to completely separate the pulling wall from the melt)
that are order $100$ times larger than $\tau_{\rm e}$.

To further investigate the role of entanglements on chain dynamics, $N_{\rm e}$
is reduced by adding a finite bending energy to the chains.
Figures \ref{fig_FvsN_ks}a and \ref{fig_FvsN_ks}b
show~$\langle F \rangle$ versus
$N_{\rm t}$ for two values of $k_{\theta}$ respectively,
at temperatures above/below $T_{\rm g}$.
This data has the same qualitative behavior on $N_{\rm t}$ as the data
in Figs.~\ref{fig_FvsN_Vs}a and \ref{fig_FvsN_Vs}b,
however the value of $N_{\rm t}$ where the onset of chain scission occurs
is smaller for chains with non-zero bending interactions.
This is consistent with the fact the $N_{\rm e}$ is reduced as chain
stiffness is increased.
The data in Fig.~\ref{fig_FvsN_ks}a and \ref{fig_FvsN_ks}b are from simulations
using $v$=$0.167 \sigma \tau^{-1}$ which is the faster of the two pulling
velocities seen in Figs.~\ref{fig_FvsN_Vs}.
However, Fig.~\ref{fig_FvsN_Vs}b shows that for $T < T_{\rm g}$ the
chain scission data is nearly independent of $v$.
Even for $T > T_{\rm g}$, the effect of finite chain stiffness on $N_{\rm e}$
(and its effect on the crossover to chain scission) is clear at the higher value
of $v$.

\subsection{Temperature dependence}

Figure \ref{fig_Wvst_N100_k0}
shows the integrated work versus time ${\cal W}(t)$,
as a function of $T$ for
$N_{\rm t}$=$100$ with $k_{\theta}$=$0.0$.
For this tethered chain length, the system is only beginning to enter
the chain scission regime for all values of $T$.
As for Fig.~\ref{fig_Wvst_T03_Ns}, the data are shown out to
sufficiently long times to show a plateau in ${\cal W}(t)$.
The data show a steady increase in
the plateau value of ${\cal W}(t)$ as $T$ increases,
contrary to what one might expect as $T$ is lowered below $T_{\rm g}$.
Since almost no breaking occurs for this value of $N_{\rm t}$, the increase
in work with decreasing $T$ can be attributed solely to an increase in the
monomeric friction coefficient.

Figures \ref{fig_Wvst_N250_ks}a and \ref{fig_Wvst_N250_ks}b
show the integrated work versus time ${\cal W}(t)$,
as a function of $T$ for
$N_{\rm t}$=$250$ with $k_{\theta}$=$0.0$ and $1.5 \epsilon$ respectively.
For $N_{\rm t}$=$250$, the system is well into the chain scission regime
for all values of $T$.
The work performed,
at early times (i.e.~$t \approx 2500 \tau$) increases monotonically as $T$ decreases
for both values of $k_{\theta}$.
However, the plateau values of
${\cal W}(t)$
show non-monotonic behavior with decreasing $T$ due to scission in the tethered chains.
Chain scission occurs throughout the pull simulation,
from early times $\propto \tau_{\rm e}$,
to times corresponding to the beginning of the ${\cal W}(t)$ plateau.
The smaller the time necessary to allow for every tethered chain
to have broken, the sooner the plateau in ${\cal W}(t)$ occurs.
The non-monotonic behavior, as well as the large jumps, in the plateau values
for ${\cal W}(t)$ suggest these effects are not due to changes exclusively
in the monomeric friction coefficient since this behavior is not present
for tethered chains that do not break (see Fig.~\ref{fig_Wvst_N100_k0}).
The plateau values for $T$=$0.6$-$1.0 \epsilon / k_{\rm B}$ are nearly identical
with a large gap between them and the higher plateau values
for all $T < 0.6 \epsilon / k_{\rm B}$.
For $k_{\theta}$=$1.5 \epsilon$, the plateau value in ${\cal W}(t)$ has a large
jump between $T$=$1.0$ and $0.6 \epsilon / k_{\rm B}$.
However, for $T < 0.6 \epsilon / k_{\rm B}$, the plateau values cluster at a {\it lower}
value of ${\cal W}(t)$.
For $N_{\rm t} > N_{\rm e}$, the effects of the increasing monomeric friction
with decreasing $T$ in addition to the longer relaxation times for entangled chains
combine to produce the unexpected, non-monotonic temperature dependence in the
${\cal W}(t)$ plateau values.

The temperature dependence of chain scission is shown explicitly in
Fig.~\ref{fig_FvsT_Ns_ks}, which shows $\langle F \rangle$ versus $T$
for the systems shown in Fig.~\ref{fig_Wvst_N100_k0} and Figs.~\ref{fig_Wvst_N250_ks}.
When $N_{\rm t}$=$100$, negligible chain scission occurs for all $T$.
For $N_{\rm t}$=$250$ and $T > T_{\rm g}$, the amount of chain scission
is constant (within errors) for both values of $k_{\theta}$.
As $T$ is lowered for $T < T_{\rm g}$, $\langle F \rangle$ decreases,
which helps explain the behavior ${\cal W}(t)$ in Fig.~\ref{fig_Wvst_N250_ks}.
More work is done because of the increased monomeric friction at low $T$,
however the increase in chain scission places
an upper limit on the total amount of work that can be done pulling the tethered
chains before they break.
For all $T$, chains with finite stiffness exhibit more scission
than fully flexible chains which is consistent with the data for $\langle F \rangle$ in
Fig.~\ref{fig_FvsN_ks} which support the fact that $N_{\rm e}$ is reduced as the
bending energy increases.

In addition to monitoring the amount of chain scission in the tethered chains,
the number of broken bonds on melt chains has also been measured.
For the fully flexible chains, $k_{\theta}$=$0.0$, a large number
of bonds in the melt are broken during a pull simulation for
$T \geq 0.6 \epsilon / k_{\rm B}$.
However, for $T < 0.6 \epsilon / k_{\rm B}$ the number of melt chain bonds
which break drops suddenly to zero.
For $k_{\theta}$=$1.5 \epsilon$, a large number
of bonds in the melt are broken for $T$=$1.0 \epsilon / k_{\rm B}$,
the number dropping to zero for $T < 1.0 \epsilon / k_{\rm B}$.
The explanation for this behavior is unclear, however there seems to be a
correlation in the work done and breaking in the melt chains.
Specifically, the large jumps in the plateau values of ${\cal W}(t)$
(Fig.~\ref{fig_Wvst_N250_ks}a and \ref{fig_Wvst_N250_ks}b)
occur when the number of broken melt bonds drops suddenly to zero.

\section{Conclusion}

In this study we present results from the first large-scale MD simulations
to study the dynamics of tethered chains in a dense polymer melt and their
effects on adhesion in a 3D, realistic polymer model.
Data is presented which shows a crossover from pure chain pullout to chain
scission which depends on the length of the tethered chains relative to
the bulk entanglement length.
For $T < T_{\rm g}$, the length of the tethered chains near this crossover
to scission is consistent with $N_{\rm e}$ (with/without bending stiffness)
for the bead-spring model used in this study.
This result agrees with experiments on reinforcing the interface between two
immiscible, glassy polymers with block copolymers.
These experiments show evidence of copolymer scission and large increases
in the work of adhesion, when the molecular weight of the copolymer is comparable
to the entanglement length.

However, the experimental data also suggest that in some cases
the increase in the work of adhesion is due to crazing.
The implication is that the copolymer block entangled in the melt transfers
sufficient stress to initiate crazing.
But crazing is not observed in our simulations with small values of the
wall interaction energy $\epsilon_{\rm wall}$
even for large $N_{\rm t}$ and $T < T_{\rm g}$.
Simulations using small values of $\epsilon_{\rm wall}$ correspond
to experiments on immiscible glassy polymers, because in both cases
the attractive forces across interface are too small to initiate crazing
in the absence of compatibilizers.
Simulations were also performed with larger values of $\epsilon_{\rm wall}$
to investigate the inverse problem: could the presence of crazing
(due to a strong interaction between the melt and the wall) alter chain scission?
These simulations showed that chain scission is virtually unaffected by the
presence of crazing (ie voids) in the melt.
The lack of crazing with small $\epsilon_{\rm wall}$ might
be because of the relatively low tethered chain areal densities simulated.
Crazing could occur at small $\epsilon_{\rm wall}$ with values of
$\Sigma$ larger than those in the mushroom regime
but still low enough to stay out of the completely phase separated regime
(see Fig.~\ref{fig_sch_WvsA}).

Future work will include developing algorithms to efficiently equilibrate
systems with tethered chains that are partially phase separated to study the
possible effects on crazing.
Also, these simulations are being extended to include realistic potentials so
that adhesion and crazing may be studied in real materials.

Sandia is a multiprogram laboratory operated by Sandia Corporation, a Lockheed 
Martin Company, for the United States Department of Energy under Contract 
DE-AC04-94AL85000.



\bibliographystyle{/usr/people/swsides/DraftPDMS/achemsol}

\begin{thebibliography}{10}

\bibitem{napper83_book}
Napper,~D.~H. \textit{Polymeric Stabilization of Colloidal Dispersions;}
  Academic: London, 1983.

\bibitem{russell89_book}
Russell,~W.~B.;\ \ Saville,~D.~A.;\ \ Schowalter,~W.~R. \textit{Colloidal
  Dispersions;} Cambridge University Press: Cambridge, 1989.

\bibitem{jklein96}
Klein,~J. \textit{Ann.\ Rev.\ Mat.\ Sci.} \textbf{1996,} \textsl{26,} 581.

\bibitem{brown93_2}
Brown,~H.~R.;\ \ Char,~K.;\ \ Deline,~V.~R.;\ \ Green,~P.~F.
  \textit{Macromolecules} \textbf{1993,} \textsl{26,} 4155.

\bibitem{riess_review75}
Riess,~G.;\ \ Jolivet,~Y.  Polyblends and Composites.   In  \textit{ACS
  Advances In Chemistry Series}, Vol.~142; Platzer,~N. A.~J.,\ \ Ed.;  American
  Chemical Society: Washington, DC, 1975.

\bibitem{deGennes80_brush}
deGennes,~P.~G. \textit{Macromolecules} \textbf{1980,} \textsl{13,} 1069.

\bibitem{aubouy95}
Aubouy,~M.;\ \ Fredrickson,~G.~H.;\ \ Pincus,~P.;\ \ Rapha{\" e}l,~E.
  \textit{Macromolecules} \textbf{1995,} \textsl{28,} 2979.

\bibitem{grest96_brushes}
Grest,~G.~S. \textit{J.\ Chem.\ Phys.} \textbf{1996,} \textsl{105,} 5532.

\bibitem{leger_review99}
L{\' e}ger,~L.;\ \ Rapha{\" e}l,~E.;\ \ Hervet,~H.  Surface-Anchored Polymer
  Chains: Their Role in Adhesion and Friction.   In  \textit{Advances In
  Polymer Science}, Vol.~138; Granick,~S.,\ \ Ed.;  Springer: Berlin, 1999.

\bibitem{kramer91}
Creton,~C.;\ \ Kramer,~E.~J.;\ \ Hadziioannou,~G. \textit{Macromolecules}
  \textbf{1991,} \textsl{24,} 1846.

\bibitem{wash_kramer94}
Washiyama,~J.;\ \ Kramer,~E.~J.;\ \ Creton,~C.~F.;\ \ Hui,~C.-Y.
  \textit{Macromolecules} \textbf{1994,} \textsl{27,} 2019.

\bibitem{kramer94_adhesion_review}
Kramer,~E.~J.;\ \ Norton,~L.~J.;\ \ Dai,~C.-A.;\ \ Sha,~Y.;\ \ Hui,~C.-Y.
  \textit{Faraday Discuss.} \textbf{1994,} \textsl{98,} 31.

\bibitem{kramer95_tether}
Norton,~L.;\ \ Smigolova,~V.;\ \ Pralle,~M.;\ \ Hubenko,~A.;\ \ Dai,~K.;\ \
  Kramer,~E.;\ \ Hahn,~S.;\ \ Berglund,~C.;\ \ DeKoven,~B.
  \textit{Macromolecules} \textbf{1995,} \textsl{28,} 1999.

\bibitem{kramer96_copoly}
Dai,~C.-A.;\ \ Kramer,~E.~J.;\ \ Washiyama,~J.;\ \ Hui,~C.-Y.
  \textit{Macromolecules} \textbf{1996,} \textsl{29,} 7536.

\bibitem{leger_tirrell95}
Deruelle,~M.;\ \ L{\' e}ger,~L.;\ \ Tirrell,~M. \textit{Macromolecules}
  \textbf{1995,} \textsl{28,} 7419.

\bibitem{stevens_fracture1_00}
Stevens,~M.~J. \textit{Macromolecules} \textbf{2001,} \textsl{34,} 1411.

\bibitem{stevens_fracture2_00}
Stevens,~M.~J. \textit{Macromolecules} \textbf{2001,} \textsl{34,} 2710.

\bibitem{robbins01}
Baljon,~A. R.~C.;\ \ Robbins,~M.~O. \textit{Macromolecules} \textbf{2001,}
  \textsl{34,} 4200.

\bibitem{robbins99}
Gersappe,~D.;\ \ Robbins,~M.~O. \textit{Euro.\ Phys.\ Lett.} \textbf{1999,}
  \textsl{48,} 150.

\bibitem{balazs96}
Pickett,~G.~T.;\ \ Jasnow,~D.;\ \ Balazs,~A.~C. \textit{Phys.\ Rev.\ Lett.}
  \textbf{1996,} \textsl{77,} 671.

\bibitem{grant99}
Sabouri-Ghomi,~M.;\ \ Ispolatov,~S.;\ \ Grant,~M. \textit{Phys.\ Rev.\ E}
  \textbf{1999,} \textsl{60,} 4460.

\bibitem{sides01_ad1}
Sides,~S.~W.;\ \ Grest,~G.~S.;\ \ Stevens,~M.~J. \textit{Phys.\ Rev.\ E.\ in
  press} \textbf{2001,} .

\bibitem{tild87}
Allen,~M.;\ \ Tildesley,~D. \textit{Computer Simulation of Liquids;} Clarendon:
  Oxford, 1987.

\bibitem{kremer_grest90}
Kremer,~K.;\ \ Grest,~G.~S. \textit{J.\ Chem.\ Phys.} \textbf{1990,}
  \textsl{92,} 5057.

\bibitem{faller00}
Faller,~R.;\ \ M\"{u}ller-Plathe,~F.;\ \ Heuer,~A. \textit{Macromolecules}
  \textbf{2000,} \textsl{33,} 6602.

\bibitem{faller00_thesis}
Faller,~R. \textit{Influence of Chain Stiffness on Structure and Dynamics of
  Polymers in the Melt,} Thesis, Max-Planck-Institut f\"{u}r Polymerforschung,
  Mainz, 2000.

\bibitem{putz00}
P\"{u}tz,~M.;\ \ Kremer,~K.;\ \ Grest,~G.~S. \textit{Euro.\ Phys.\ Lett.}
  \textbf{2000,} \textsl{49,} 735 The simulations in this paper used a purely
  repulsive interaction between non-bonded monomers and a FENE interaction
  between bonded monomers. The $N_{\rm e}$ for this model is expected to be
  very similar to that of the model used in the present study.

\bibitem{faller01}
Faller,~R. \textit{to be published} \textbf{2001,} .

\bibitem{plimpton95}
Plimpton,~S. \textit{J.\ Comp.\ Phys.} \textbf{1995,} \textsl{117,} 1.

\bibitem{rast3D}
Merritt,~E.~A.;\ \ Bacon,~D.~J. \textit{Meth.\ Enzymol.} \textbf{1997,}
  \textsl{277,} 505.

\bibitem{kremer_grest_review}
Kremer,~K.;\ \ Grest,~G.~S.  Entanglement Effects In Polymer Melts and
  Networks.   In  \textit{Monte Carlo and Molecular Dynamics Simulations in
  Polymer Science}; Binder,~K.,\ \ Ed.;  Oxford University Press: New York,
  1995.

\end{thebibliography}

\providecommand{\refin}[1]{\\ \textbf{Referenced in:} #1}



\begin{figure}[tbh]
\caption
{
\label{fig_sch_WvsA}
Schematic of the increased work of adhesion $W^*$ for an elastomeric system
versus areal density of tethered chains $\Sigma$.
(Region I) Mushroom regime,
(Region II) Partially overlapped (wet brush) regime,
(Region III) Phase separated regime.
$W^*$ shows a similar dependence on $N_{\rm t}$.
}
\end{figure}


\begin{figure}[tbh]
\caption
{
\label{fig_ubreak}
Comparison of the breaking potential $U_{\rm b}(r)$
and the unbreakable FENE potential $U_{\rm nb}(r)$.
}
\end{figure}


\begin{figure}[tbh]
\caption
{
\label{fig_sch_bend}
Schematic for bending potential.
The three labeled monomers define the triplet used in calculating the
bending force associated with the angle $\theta$.
}
\end{figure}

\begin{figure}
\caption
{
\label{fig_pic_n100}
(color) Chain configurations at two times during a tensile
pull simulation for $N_{\rm t}$=$100$
with $T$=$0.3 \epsilon / k_{\rm B}$,
pull velocity $v$=$0.167 \sigma \tau^{-1}$ and $k_{\theta}$=$0.0$.
Elapsed times shown are
(a) $300 \tau$ and
(b) $600 \tau$.
The red monomers belong to tethered chains and blue monomers
belong to melt chains.
Green monomers belong to sections of tethered
chains which are no longer attached to the bottom substrate.
Raster3D \protect\cite{rast3D} is used to render the images.
}
\end{figure}

\begin{figure}
\caption
{\label{fig_pic_n250}
(color) Chain configurations at three times during a tensile
pull simulation for
$N_{\rm t}$=$250$ with
$T$=$0.3 \epsilon / k_{\rm B}$,
pull velocity $v$=$0.167 \sigma \tau^{-1}$ and $k_{\theta}$=$0.0$.
Elapsed times shown are
(a) $300 \tau$,
(b) $900 \tau$ and
(c) $1200\tau$.
Coloring scheme is the same as for Fig.~\protect{\ref{fig_pic_n100}}.
}
\end{figure}

\begin{figure}[htb]
\caption
{
\label{fig_Wvst_T03_Ns}
Total integrated work ${\cal W}(t)$ vs.~time at
different tethered chain lengths $N_{\rm t}$ for
$T$=$0.3 \epsilon / k_{\rm B}$,
$v$=$0.0167 \sigma \tau^{-1}$ and
$k_{\theta}$=$0.0$.
}
\end{figure}

\begin{figure}[htb]
\caption
{\label{fig_FvsN_Vs}
Average fractional length of remaining tethered chains~$\langle F \rangle$
vs.~initial tethered chain length $N_{\rm t}$ for
(a) $T$=$1.0 \epsilon / k_{\rm B}$,
(b) $T$=$0.3 \epsilon / k_{\rm B}$.
The symbols denote
$v$=$0.167  \sigma \tau^{-1}$ (filled square) and
$v$=$0.0167 \sigma \tau^{-1}$ (filled triangle).
The open squares show similar results for a crosslinked network
with $\approx 18$ crosslinks per melt chain for
$v$=$0.0167 \sigma \tau^{-1}$.
For all data $k_{\theta}$=$0.0$.
}
\end{figure}

\begin{figure}[htb]
\caption
{
\label{fig_FvsV_Ts}
Average fractional length of remaining tethered chains~$\langle F \rangle$
vs.~tensile pull velocity $v$ for $N_{\rm t}$=$250$
at
$T$=$1.0 \epsilon / k_{\rm B}$ for the
crosslinked (open triangles)
and
uncrosslinked (solid triangles) system
and at
$T$=$0.3 \epsilon / k_{\rm B}$
for the uncrosslinked system (solid squares).
For all data $k_{\theta}$=$0.0$.
}
\end{figure}

\begin{figure}[htb]
\caption
{\label{fig_FvsN_ks}
Average fractional length of remaining tethered chains~$\langle F \rangle$
vs.~initial tethered chain length $N_{\rm t}$ for
(a) $T$=$1.0 \epsilon / k_{\rm B}$,
(b) $T$=$0.3 \epsilon / k_{\rm B}$.
The symbols denote
$k_{\theta}$=$0.0         $ (filled square) and
$k_{\theta}$=$1.5 \epsilon$ (filled triangle).
For all data $v$=$0.167  \sigma \tau^{-1}$.
}
\end{figure}

\begin{figure}[htb]
\caption
{
\label{fig_Wvst_N100_k0}
Total integrated work ${\cal W}(t)$ vs.~time as a function of
temperature for
$N_{\rm t}$=$100$,
$v$=$0.0167 \sigma \tau^{-1}$ and
$k_{\theta}$=$0.0$.
}
\end{figure}

\begin{figure}[htb]
\caption
{
\label{fig_Wvst_N250_ks}
Total integrated work ${\cal W}(t)$ vs.~time as a function of
temperature for $N_{\rm t}$=$250$ and
$v$=$0.0167 \sigma \tau^{-1}$.
(a) $k_{\theta}$=$0.0$
(b) $k_{\theta}$=$1.5 \epsilon$.
}
\end{figure}

\begin{figure}
\caption
{
\label{fig_FvsT_Ns_ks}
Average fractional length of remaining
tethered chains~$\langle F \rangle$ vs.~temperature $T$ for
$v$=$0.0167  \sigma \tau^{-1}$.
The symbols denote
$N_{\rm t}$=$100$, $k_{\theta}$=$0.0$ (filled square),
$N_{\rm t}$=$250$, $k_{\theta}$=$0.0$ (filled triangle)
and
$N_{\rm t}$=$250$, $k_{\theta}$=$1.5 \epsilon$ (open triangle).
}
\end{figure}

\end{document}